\documentclass[letterpaper,10pt]{article}

\usepackage{amssymb}
\usepackage{amsmath}

\newcommand{\dd}{\textrm{d}}
\newcommand{\im}{{\mathbb{I}}{\mathrm{m}}}

\author{A. L\'opez-Ortega\thanks{alopezo@ipn.mx} \\
Centro de Investigaci\'on en Ciencia Aplicada y Tecnolog\'{\i}a Avanzada. \\ 
Unidad Legaria. Instituto Polit\'ecnico Nacional. \\
Calzada Legaria \# 694. Colonia Irrigaci\'on. Delegaci\'on Miguel Hidalgo. \\
M\'exico, D.\ F., M\'exico. \\
C.\ P.\  11500  
}

\title{On the Time Times Temperature Bound}

\begin{document}

\maketitle

\begin{abstract}

Recently Hod proposes a lower bound on the relaxation time of a perturbed thermodynamic system. For gravitational systems this bound transforms into a condition on the fundamental quasinormal frequency. We test the bound in some spacetimes whose quasinormal frequencies are calculated exactly, as the three-dimensional BTZ black hole, the $D$-dimensional de Sitter spacetime, and the $D$-dimensional Nariai spacetime. We find that for some of these spacetimes their fundamental quasinormal frequencies do not satisfy the bound proposed by Hod.

\end{abstract}


\section{Introduction}
\label{section 1}

The quasinormal modes (QNM) characterize the response of a black hole to external perturbations \cite{Kokkotas:1999bd,Berti:2009kk}. Their complex frequencies, the so called quasinormal frequencies (QNF), determine the oscillation frequency and the decay time of the field that moves in the black hole. Furthermore the QNF depend on the physical parameters of the black hole, thus the QNM provide us with a useful tool to measure its physical properties. Some time ago the applications in astrophysics motivated the study of the QNM \cite{Kokkotas:1999bd,Berti:2009kk}. Recently the QNM have found applications in other lines of research (see Refs.\ \cite{Horowitz:1999jd}--\cite{Denef:2009kn} for some examples).

For a spacetime its fundamental mode, that is, the least damped QNM, determines the decay time of the perturbation field when it propagates in the background. Furthermore in a spacetime the decay time of the perturbation is equal to
\begin{equation}
 \tau = \frac{1}{\omega_I},
\end{equation} 
where $\omega_I$ stands for the absolute value of the imaginary part of the fundamental QNF $\omega$ \cite{Kokkotas:1999bd,Berti:2009kk,Hod:2006jw}.

Analyzing how a perturbed thermodynamic system returns to an equilibrium state, and based on information theory and thermodynamics, Hod finds that the relaxation time $\tau$ of a perturbed thermodynamic system satisfies \cite{Hod:2006jw}
\begin{equation} \label{eq: Hod bound one}
 \tau \geq \tau_{min} = \frac{\hbar}{\pi T},
\end{equation} 
where $\hbar$ denotes the reduced Planck constant, and $T$ stands for the temperature of the system. Thus, according to Hod, for a perturbed thermodynamic system there is (at least) one perturbation mode whose time of relaxation is $\tau_{min}$ or larger. The lower bound (\ref{eq: Hod bound one}) is named Time Times Temperature bound by Hod \cite{Hod:2006jw}, (\textit{TTT bound} in what follows).

At least other two derivations of the TTT bound have been presented \cite{Ropotenko:2007jm,Pesci:2008zv}. Ropotenko deduces the TTT bound from a well known condition that determines whether a perturbed quantity relaxes in a classical way \cite{Ropotenko:2007jm}. Using the existence of a minimal length scale for the size of a thermodynamic system in order to have a significant notion of statistical entropy, Pesci shows that, at least for some systems, the TTT bound is valid \cite{Pesci:2008zv}.

Furthermore Hod notes that strong self-gravity systems are appropriate to test the TTT bound \cite{Hod:2006jw}. For these systems he shows that lower bound (\ref{eq: Hod bound one}) transforms into an upper bound on the absolute value of the imaginary part of the fundamental QNF $\omega_I$ (see formula (4) in Ref.\ \cite{Hod:2006jw})
\begin{equation} \label{eq: Hod bound}
 \omega_I \leq \frac{\pi T_H}{\hbar} ,
\end{equation} 
where $T_H$ stands for Hawking's temperature of the spacetime. Defining the quantity $\mathbb{H}$ by
\begin{equation} \label{eq: H quantity}
 \mathbb{H} = \frac{\hbar \omega_I}{\pi T_H},
\end{equation} 
we find that upper bound (\ref{eq: Hod bound}) becomes $\mathbb{H} \leq 1$. Notice that if upper bound (\ref{eq: Hod bound}) is not valid, that is for the fundamental mode $\mathbb{H} > 1$, then the decay time of the fundamental QNF for the system is less than $\tau_{min}$ and the inequality (\ref{eq: Hod bound one}) is not satisfied.

In Refs.\ \cite{Hod:2006jw}, \cite{Hod:2007tb}--\cite{Hod:2008se}, bound (\ref{eq: Hod bound}) is tested in the black holes: $D$-dimensional Schwarzschild, $D$-dimensional Schwarzschild de Sitter, large and small Schwarzschild anti de Sitter, Kerr, Kerr-Newman, extreme Kerr, and near extreme Kerr-Newman (see also Ref.\ \cite{Gruzinov:2007ai}). For these spacetimes Hod finds that upper bound (\ref{eq: Hod bound}) is satisfied, and in particular he finds that the extremal Kerr black hole saturates it \cite{Hod:2007tb,Hod:2008zz}, (moreover it is expected that in the extremal limit the Kerr-Newman and Schwarzschild de Sitter black holes saturate upper bound (\ref{eq: Hod bound})) \cite{Hod:2006jw,Hod:2008se}. Notice that for these black holes the QNF are computed numerically or calculated analytically in a particular limit. For example, for the  Kerr black hole the QNF are calculated analytically in the extremal limit \cite{Hod:2007tb,Hod:2008zz}.

According to Hod: \textit{``It is of great interest to check the validity of the upper bound, Eq.\ (4), for other black hole spacetimes.''} (Page 3 of Ref.\ \cite{Hod:2006jw} and we note that in the previous sentence ``Eq.\ (4)'' refers to formula (4) of Ref.\ \cite{Hod:2006jw} and it is equal to our formula (\ref{eq: Hod bound}).) Therefore we believe that it is pertinent to test the TTT bound in spacetimes whose QNF are calculated exactly. 

As far as we know the QNF of several fields are calculated exactly for: 
\begin{enumerate}

\item three-dimensional charged and rotating black holes of the Einstein-Maxwell-dilaton with cosmological constant theory \cite{Fernando:2003ai,LopezOrtega:2005ep,Fernando:2008hb,Fernando:2009tv},

\item five-dimensional dilatonic black hole \cite{Becar:2007hu,LopezOrtega:2009zx},

\item two-dimensional dilatonic black hole \cite{Becar:2007hu,LopezOrtega:2009zx},

\item four-dimensional BTZ black string \cite{Liu:2008ds},

\item Chern-Simons black holes \cite{Gonzalez:2010vv},

\item Warped $AdS_3$ black holes (spacelike and null stretched) \cite{Chen:2009hg,Chen:2009rf},

\item $D$-dimensional massless topological black hole ($D \geq 4$) \cite{Aros:2002te,Birmingham:2006zx,LopezOrtega:2010,LopezOrtega:2007vu}, (see Ref.\ \cite{Oliva:2010xn} for an application and extension of some results for the MTBH to other black holes whose metrics are similar),

\item three-dimensional static and spinning BTZ black holes \cite{Birmingham:2001pj,Cardoso:2001hn,Birmingham:2001hc,Decanini:2009dn,Sachs:2008gt,Crisostomo:2004hj},

\item $D$-dimensional de Sitter spacetime ($D \geq 4$) \cite{Du:2004jt,Natario:2004jd,Choudhury:2003wd,Lopez-Ortega:2006my,Zelnikov:2008rg,Lopez-Ortega:2006ig,LopezOrtega:2007sr},

\item $D$-dimensional Nariai spacetime ($D \geq 4$) \cite{LopezOrtega:2007vu,Vanzo:2004fy,LopezOrtega:2009qc}.

\end{enumerate}

We point out that for the three-dimensional charged and rotating black holes of the Einstein-Maxwell-dilaton with cosmological constant theory, the five-dimensional dilatonic black hole, and the two-dimensional dilatonic black hole, for the fields whose frequencies are known exactly \cite{Fernando:2003ai,LopezOrtega:2005ep,Fernando:2008hb,Fernando:2009tv,Becar:2007hu,LopezOrtega:2009zx}, the fundamental mode may be unstable, and as a consequence in what follows these three black holes are not studied. We note that the unstable behavior of the fundamental mode depends on the parameters of the black hole and the field. Sometimes this instability does not occur. See Refs.\ \cite{Fernando:2003ai}--\cite{LopezOrtega:2009zx} for more details.

Furthermore in Ref.\ \cite{Liu:2008ds} for the four-dimensional BTZ black string are computed exactly the QNF for the massless Dirac field and the scalar type gravitational perturbation. Notice that in Ref.\ \cite{Liu:2008ds} some restrictions are imposed on the modes whose QNF are computed exactly. For the Chern-Simons black holes only the QNF of the Klein-Gordon field are calculated exactly \cite{Gonzalez:2010vv}. Here we do not test upper bound (\ref{eq: Hod bound}) in these two spacetimes because it is necessary to know the spectrum of the QNF in more detail.

For the warped $AdS_3$ black holes the QNF of the Klein-Gordon, Dirac, and massive vector fields are computed exactly \cite{Chen:2009hg,Chen:2009rf}. In what follows we do not analyze the warped $AdS_3$ black holes. We have two reasons: i) Notice that in Refs.\ \cite{Chen:2009hg}, \cite{Chen:2009rf} are presented two sets of quantities that are called the QNF of the warped $AdS_3$ black holes. The first set of QNF is calculated with the usual procedure, that is, solving the equations of motion for the field and then imposing the boundary conditions of the QNM. The second set of QNF is obtained from the first by means a change of coordinates in order to match the predictions of the warped AdS-CFT correspondence \cite{Chen:2009hg,Chen:2009rf}.  We do not know what set of QNF must be used to test the TTT bound. ii) The warped $AdS_3$ black holes are solutions of the topological massive gravity and this theory there is a propagating graviton, whose QNF are not know at present time (as far as we know). We believe that the massive graviton is the appropriate field to test the TTT bound.

In Ref.\ \cite{LopezOrtega:2010}, upper bound (\ref{eq: Hod bound}) is tested in $D$-dimensional ($D \geq 4$) massless topological black hole (MTBH in what follows). In that reference is asserted that the TTT bound is not satisfied in MTBH. Although we believe that the conclusion obtained in Ref.\ \cite{LopezOrtega:2010} is true, it is necessary to clarify some points that in its exposition are confusing.

The metric of the $D$-dimensional MTBH reads
\begin{equation} \label{eq: line element MTBH}
 \dd s^2 = - \left(-1 + \frac{r^2}{L^2} \right) \dd t^2 + \left(-1 + \frac{r^2}{L^2} \right)^{-1} \dd r^2  + r^2 \dd \Sigma_{D-2}^2,
\end{equation}  
where $L$ is related to the negative cosmological constant $\Lambda$ by
\begin{equation}
 L^2 = - \frac{(D-1)(D-2)}{2 \Lambda} ,
\end{equation}
and $\dd \Sigma^2_{D-2}$ stands for the line element of a $(D-2)$-dimensional compact Einstein space of negative curvature \cite{Vanzo:1997gw,Mann:1997iz,Birmingham:1998nr,Lemos:1994xp}.

For the MTBH the QNF for the three types of the gravitational perturbations  are \cite{Birmingham:2006zx,LopezOrtega:2010}
\begin{equation} \label{eq: bosons MTBH}
 \omega =\pm \frac{\xi}{L} - \frac{2i}{L}\left( n + \frac{\mathbb{A}}{4} \right) ,
\end{equation} 
where $\xi$ is related to the eigenvalues of the Laplacian on the manifold $\Sigma_{D-2}$, $n=0,1,2,\dots$, is the mode number, and
 \begin{align} \label{eq: A values MTBH}
 \mathbb{A}=  & \left\{ \begin{array}{l}  D-1 \qquad \, \, \, \, \, \, \textrm{for the vector type gravitational perturbations,} \\ 
 |D-5|+2  \, \, \, \, \textrm{for the scalar type gravitational perturbations,} \\
 D+1  \, \,\, \,\, \, \qquad \textrm{for the tensor type gravitational perturbations ($D\geq5$),} \\
 \end{array} \right.
\end{align}   
(see formulas (46) of Ref.\ \cite{Birmingham:2006zx}).

Taking into account that Hawking's temperature of the MTBH is \cite{Birmingham:2006zx}
\begin{equation} 
 T_{MTBH} = \frac{\hbar}{2 \pi L},
\end{equation} 
in Ref.\ \cite{LopezOrtega:2010} it is shown that for the three types of gravitational perturbations
\begin{align} \label{eq: H values MTBH}
 \frac{\hbar |\im(\omega)|_{n=0}}{\pi T_{MTBH}} = \mathbb{A},
\end{align} 
that is, in Ref.\ \cite{LopezOrtega:2010} it is calculated ratio (\ref{eq: H values MTBH}) of the modes $n=0$ for the three types of gravitational perturbations. From this result in Ref.\ \cite{LopezOrtega:2010} it is stated that the TTT bound is not satisfied in the MTBH. Nevertheless, to test the TTT bound it is necessary to find the least damped mode and not only the modes with $n=0$ for each type of the gravitational perturbations. Thus the implicit assumption of Ref.\ \cite{LopezOrtega:2010} $\mathbb{H} = \mathbb{A}$ is not valid. For the $D$-dimensional MTBH with $D \geq 5$ the fundamental mode corresponds to scalar type gravitational perturbations with $n=0$. Hence for $D \geq 5$  in the $D$-dimensional MTBH we get that
\begin{equation} 
\mathbb{H}=|D-5|+2 .
\end{equation} 
For $D \geq 5$ notice that if the dimension of the MTBH increases, then the quantity $\mathbb{H}$ increases and the decay time of the fundamental QNM decreases. 

For the four-dimensional MTBH, from formulas (\ref{eq: A values MTBH}) and (\ref{eq: H values MTBH}), we find that for the fundamental QNF of the vector or scalar type gravitational perturbations
\begin{equation}
 \mathbb{H} = 3. 
\end{equation} 
Thus even in four dimensions the fundamental QNF of the MTBH does not satisfy upper bound (\ref{eq: Hod bound}). Hence we found that for the gravitational perturbations of the $D$-dimensional MTBH its fundamental QNF does not satisfy upper bound (\ref{eq: Hod bound}) and therefore in the $D$-dimensional MTBH the decay time of the fundamental mode is less than $\tau_{min}$. This conclusion is identical to that already presented in Ref.\ \cite{LopezOrtega:2010}, although the conclusion of the previous reference is based on some unjustified assumptions. 

Notice that we do not extend the results for the MTBH to the black holes studied by Oliva and Troncoso in Ref.\ \cite{Oliva:2010xn} because in these spacetimes the QNF of the gravitational perturbations have not been calculated (as far as we know).

Our aim is to extend the previous results on the MTBH and test upper bound (\ref{eq: Hod bound}) in three-dimensional BTZ black hole, $D$-dimensional de Sitter spacetime, and $D$-dimensional Nariai spacetime ($D \geq 4$). Although the de Sitter spacetime and the Nariai spacetime are not black holes, we believe that it is relevant to test bound (\ref{eq: Hod bound}) in these backgrounds.

This paper is organized as follows. In Sec.\ \ref{section 2} we use the fundamental QNF of the three-dimensional spinning BTZ black hole to test upper bound (\ref{eq: Hod bound}). In Sec.\ \ref{section 3} we study a similar question in the $D$-dimensional de Sitter spacetime. In Sec.\ \ref{section 4} we use the fundamental QNF of the $D$-dimensional Nariai spacetime to test upper bound (\ref{eq: Hod bound}). Finally in Sec.\ \ref{section 5} we discuss the results obtained.

\section{Three-dimensional BTZ black hole}
\label{section 2}

A relevant solution of the three-dimensional Einstein's equations with negative cosmological constant $\Lambda$ is the spinning BTZ black hole. In units where $8G = 1$, the metric of this black hole is \cite{Banados:1992wn,Banados:1992gq,Carlip:1995qv} 
\begin{align} \label{eq: BTZ metric}
 \dd s^2 &= -\left( -M + \frac{r^2}{l^2} + \frac{J^2}{4 r^2}\right) \dd t^2  + \left( -M + \frac{r^2}{l^2} + \frac{J^2}{4 r^2}\right)^{-1} \dd r^2 + r^2 \left( \dd \phi - \frac{J}{2 r^2} \dd t \right)^2 ,
\end{align} 
where the quantity $l$ is related to the three-dimensional negative cosmological constant by $\Lambda = -1/l^2$, $M$ and $J$ denote the mass and the angular momentum of the BTZ black hole. We write these two quantities as \cite{Banados:1992wn,Banados:1992gq,Carlip:1995qv} 
\begin{equation} \label{eq: BTZ M J}
 M = \frac{r_+^2 + r_-^2}{l^2}, \qquad \qquad \qquad J = \frac{2 r_+ r_-}{l},
\end{equation} 
with $r_+$ denoting the radius of the event horizon and $r_-$ the radius of the inner horizon. The radii $r_+$ and  $r_-$ are equal to
\begin{equation} \label{eq: BTZ radii}
 r_{\pm}^2 = \frac{M l^2}{2}\left[1 \pm \left(1-\frac{J^2}{M^2 l^2} \right)^{1/2} \right]. 
\end{equation} 

In the spinning BTZ black hole the QNF of the Klein-Gordon field of mass $\mu$ are calculated exactly in Ref.\ \cite{Birmingham:2001hc} (see also Ref.\ \cite{Birmingham:2001pj})
\begin{align}  \label{eq: QNF BTZ}
 \omega_L &= \frac{m}{l} - 2 i \left(\frac{r_+ - r_-}{l^2} \right) \left(n + \frac{1}{2} + \frac{1}{2}\sqrt{1 + \mu^2 l^2} \right),  \\
\omega_R &= - \frac{m}{l} - 2 i \left(\frac{r_+ + r_-}{l^2} \right) \left(n + \frac{1}{2} + \frac{1}{2}\sqrt{1 + \mu^2 l^2} \right), \nonumber
\end{align}
where $m$ is the azimuthal number, and $n=0,1,2,\dots$, is the mode number. See also Refs.\ \cite{Birmingham:2001pj}--\cite{Crisostomo:2004hj} for the QNF of other fields in the spinning and static BTZ black holes.\footnote{Notice that Birmingham writes the Klein-Gordon equation in the form (see formula (5) in Ref.\ \cite{Birmingham:2001hc})
\begin{equation}
 \left(\square^2 - \frac{\mu}{l^2}\right)\Phi = 0,  \nonumber
\end{equation} 
where $\square^2$ stands for the three-dimensional d'Alembertian and $\Phi$ denotes the Klein-Gordon field. In this paper we write the massive Klein-Gordon equation in the form
\begin{equation}
 (\square^2 - \mu^2)\Phi = 0.  \nonumber
\end{equation} }

Taking into account that Hawking's temperature of the BTZ black hole is \cite{Carlip:1995qv}
\begin{equation}
 T_{BTZ} = \frac{\hbar (r_+^2-r_-^2)}{2\pi l^2 r_+},
\end{equation} 
and since $|\im(\omega_L)|<|\im(\omega_R)|$ for the same values of $n$ and $\mu$, we find $\omega_I = |\im(\omega_L)|_{n=0}$. Thus for the massive Klein-Gordon field moving in the BTZ black hole we find that the quantity $\mathbb{H}$ is equal to 
\begin{equation}
 \mathbb{H} = (2 + 2 \sqrt{1 + \mu^2 l^2}) \frac{r_+}{r_+ + r_-} \geq 2 .
\end{equation} 
Hence the quantity $\mathbb{H}$ is greater than 1. Thus for the massive Klein-Gordon field propagating in the spinning BTZ black hole we find that its fundamental QNF does not satisfy upper bound (\ref{eq: Hod bound}). Therefore in this spacetime we get that the decay time of the fundamental QNM is less than $\tau_{min}$. We also note that for the QNF $\omega_R$ with $n=0$
\begin{equation}
 \mathbb{H}_R = \frac{\hbar |\im(\omega_R)|_{n=0}}{\pi T_{BTZ}}=(2 + 2 \sqrt{1 + \mu^2 l^2}) \frac{r_+}{r_+ - r_-} \geq 4 .
\end{equation} 

According to the AdS/CFT correspondence the QNF of the BTZ black hole correspond to the poles of the retarded Green function for the dual conformal field theory (CFT) \cite{Birmingham:2001pj}. Furthermore the imaginary parts of the QNF are interpreted as the time scale for the thermalization of small perturbations for the dual CFT. Thus for the dual CFT of the BTZ black hole, upper bound (\ref{eq: Hod bound}) predicts that the retarded Green function has at least one pole whose absolute value of its imaginary part is less than $\pi T_{BTZ} / \hbar $. In Ref.\ \cite{Birmingham:2001pj} for the dual CFT theory of the BTZ black hole the poles of the retarded Green function are calculated exactly, and an analysis shows that there are no poles whose absolute values of their imaginary parts be less than $\pi T_{BTZ} / \hbar $.

In string theory it is interesting to know the range of energy and momentum for which the semiclassical decay rate of the BTZ black hole can be reproduced by a dual CFT at finite temperature. This problem is studied in Refs.\ \cite{Birmingham:1997rj}--\cite{Dasgupta:1998jg}. In the previous papers the so-called left and right temperatures of the BTZ black hole are defined and used (see also Refs.\ \cite{Birmingham:2001pj}, \cite{Birmingham:2001hc} on the QNM of the spinning BTZ black hole)
\begin{equation} \label{eq: left and right temperature}
 T_L = \frac{\hbar (r_+ - r_-) }{2 \pi l^2}, \qquad \qquad  T_R = \frac{\hbar (r_++r_-) }{2 \pi l^2}.
\end{equation}  

Since in the spinning BTZ black hole upper bound (\ref{eq: Hod bound}) is not satisfied, we also investigate whether this bound is fulfilled when in the definition of the quantity $\mathbb{H}$ we use the left temperature (for the QNF $\omega_L$) and the right temperature (for the QNF $\omega_R$) instead of Hawking's temperature. 

Taking into account formulas (\ref{eq: QNF BTZ}) and (\ref{eq: left and right temperature}) we find
\begin{align}
 \frac{\hbar |\im(\omega_L)|_{n=0}}{\pi T_L} =\frac{\hbar |\im(\omega_R)|_{n=0}}{\pi T_R} = 2 + 2 \sqrt{1 + \mu^2 l^2} \geq 4. 
\end{align} 
Although the substitution of Hawking's temperature by the left and right temperatures, we obtain that in the spinning BTZ black hole the fundamental QNF of the massive Klein-Gordon field do not satisfy upper bound (\ref{eq: Hod bound}).

We notice that the black hole solutions of Einstein's equations with negative cosmological constant also are solutions to the equations of motion for topologically massive gravity \cite{Sachs:2008gt}. Furthermore in topologically massive gravity there is a propagating graviton of mass $\mu$. As in Ref.\ \cite{Sachs:2008gt}, in what follows we assume that $\mu$ is a rational number and $|\mu| \geq 1$.

Considering the non-rotating BTZ  black hole (i.\ e.\  $J=r_-=0$) as a solution of topologically massive gravity, Sachs and Solodukhin calculate exactly the QNF of its tensor perturbations \cite{Sachs:2008gt}. From the results of Ref.\ \cite{Sachs:2008gt} we note that for each sign of $\mu$ the corresponding QNM have a definite chirality.

In Ref.\ \cite{Sachs:2008gt} Sachs and Solodukhin  find that in the non-rotating BTZ black hole for $\mu < -1$ the QNF of the tensor perturbations are\footnote{As in Ref.\ \cite{Sachs:2008gt}, in the rest of this section we use units where $l=1$, $r_+=1$.} (see formulas (3.32) and (3.38) of Ref.\ \cite{Sachs:2008gt})
\begin{equation} \label{eq: BTZ tensor L}
 \omega_R = m -2i(h_R+n),
\end{equation} 
whereas $\mu > 1$ the QNF are
\begin{equation} \label{eq: BTZ tensor R}
 \omega_L = -m -2i(h_L+n),
\end{equation} 
where
\begin{equation}
 h_R=h_L=\frac{|\mu|}{2} -\frac{1}{2}.
\end{equation} 

For tensor perturbations moving in the static BTZ black hole the least damped mode correspond to $n=0$ for the QNF $\omega_R$ (\ref{eq: BTZ tensor L}) or $\omega_L$ (\ref{eq: BTZ tensor R}). Thus for the fundamental QNF of the tensor perturbations we get
\begin{equation}
 \mathbb{H} = |\mu| -1.
\end{equation} 
From this expression we obtain that $\mathbb{H} > 1$ for $|\mu| > 2$. Hence in topologically massive gravity for the non-rotating BTZ black hole we find that when the mass of the graviton satisfies $|\mu| > 2$ the fundamental mode of the tensor perturbations do not satisfy upper bound (\ref{eq: Hod bound}) and hence its decay time is less than $\tau_{min}$.

As we previously mentioned, Hod shows that the fundamental QNF of the extremal Kerr black hole saturates bound (\ref{eq: Hod bound}) \cite{Hod:2007tb,Hod:2008zz}.\footnote{A photon gas is another example of a system that saturates the lower bound (\ref{eq: Hod bound one}) \cite{Pesci:2008zv}.} For the extreme BTZ black hole the QNF are not well defined, in this spacetime only exist normal modes, and the extreme BTZ black hole behaves as a non-dissipative system \cite{Crisostomo:2004hj}. Therefore in this spacetime we do not test upper bound (\ref{eq: Hod bound}).

\section{$D$-dimensional de Sitter spacetime}
\label{section 3}

In this section we study the $D$-dimensional de Sitter spacetime. In static coordinates its line element reads \cite{Spradlin:2001pw,Kim:2002uz}
\begin{equation} \label{e: static de Sitter}
{\rm d} s^2 = - \left(1 - \frac{r^2}{L^2} \right) {\rm d} t^2 + \left(1 - \frac{r^2}{L^2} \right)^{-1} {\rm d}r^2  + r^2 {\rm d} \Omega_{D-2}^2,
\end{equation} 
with ${\rm d} \Omega_{D-2}^2$ denoting the line element of a $(D-2)$-dimensional sphere, and $L^2$ is related to the positive cosmological constant $\Lambda$ by
\begin{equation}
 L^2 = \frac{(D-1)(D-2)}{2 \Lambda} .
\end{equation} 
As is well known the de Sitter spacetime has a cosmological horizon at $r=L$. In what follows the spacetime dimension fulfills $D\geq4$. 

For the $D$-dimensional de Sitter spacetime the QNF of the gravitational perturbations are equal to \cite{Natario:2004jd,Lopez-Ortega:2006my}
\begin{align} \label{eq: QNF dS gravitational}
 \omega_1  = - \frac{i}{L}(l + D -1 - q + 2 n), \qquad \qquad \omega_2  = - \frac{i}{L}(l + q + 2n),  
\end{align} 
where $q=0$ for tensor type ($D\geq5$), $q=1$ for vector type, $q=2$ for scalar type gravitational perturbations, $n$ is the mode number ($n = 0, 1,\dots $), and $l$ is the orbital quantum number, with $l\geq 2$ for the gravitational perturbations (see formulas (37) and (38) of Ref.\ \cite{Lopez-Ortega:2006my}).\footnote{In contrast to some claims \cite{Natario:2004jd,Choudhury:2003wd}, we believe that for the $D$-dimensional de Sitter spacetime the QNF of the gravitational perturbations are well defined in even and odd dimensions \cite{Lopez-Ortega:2006my}. } 

Thus in $D$-dimensional de Sitter spacetime from formulas (\ref{eq: QNF dS gravitational}) we find
 \begin{align} \label{eq: imaginary part dS}
 |\im(\omega_{1})|  = \frac{1}{L}(l + D -1 - q +2n), \qquad \qquad |\im(\omega_{2})| = \frac{1}{L}(l + q +2n),  
\end{align} 
and for $D \geq 5$ we obtain that $|\im(\omega_2)| \leq |\im(\omega_1)|$ for the same values of mode number $n$, angular momentum number $l$, and spacetime dimension $D$. From these expressions we find that in de Sitter spacetime with $D \geq 5$ the fundamental QNF corresponds to $\omega_2$ for tensor type perturbations ($q=0$) with $l=2$.

Taking into account that Hawking's temperature of the de Sitter cosmological horizon is \cite{Gibbons:1977mu}
\begin{equation} \label{e: temperature dS}
 T_{dS} = \frac{\hbar}{2 \pi L},
\end{equation} 
we find that in $D$-dimensional de Sitter spacetime with $D \geq 5$ the quantity $\mathbb{H}$ is equal to
\begin{align} \label{eq. H dS} 
 \mathbb{H} & = 4.
\end{align}  

From formulas (\ref{eq: QNF dS gravitational}) we obtain that in the four-dimensional de Sitter spacetime the fundamental QNF corresponds to the scalar or vector type gravitational perturbations and it reads
\begin{equation}
 \omega = -\frac{i}{L}(l+1),
\end{equation} 
with $l=2$. Taking into account the expression for Hawking's temperature (\ref{e: temperature dS}), in four dimensions for the quantity $\mathbb{H}$ we find the result 
\begin{equation} \label{eq. H dS 4} 
 \mathbb{H} = 6 .
\end{equation} 

We notice that for $D \geq 5$ in de Sitter spacetime the least damped mode corresponds to tensor type gravitational perturbations, whereas in $D=4$ the least damped mode corresponds to scalar or vector type gravitational perturbations. As a consequence in de Sitter spacetime for $D = 4$ and $D \geq 5$ we get different values of the quantity $\mathbb{H}$ .

Therefore from formulas (\ref{eq. H dS}) and (\ref{eq. H dS 4}) for the $D$-dimensional de Sitter spacetime ($D \geq 4$) we obtain that the fundamental QNF of the gravitational perturbations does not satisfy upper bound (\ref{eq: Hod bound}). Furthermore in de Sitter spacetime the decay time of the fundamental QNM is less than $\tau_{min}$ and lower bound (\ref{eq: Hod bound one}) is not valid.

According to Hod \cite{Hod:2006jw}, the thermodynamic systems whose temperature depends on the inverse of its characteristic length, it is more probable that saturate upper bound (\ref{eq: Hod bound}). We notice that Hawking's temperature of the $D$-dimensional de Sitter horizon depends on the inverse of the cosmological horizon radius, which is the characteristic length of the de Sitter spacetime, but we find that for the $D$-dimensional de Sitter spacetime the fundamental QNF does not satisfy upper bound (\ref{eq: Hod bound}).

\section{$D$-dimensional Nariai spacetime }
\label{section 4}

The metric of the $D$-di\-men\-sion\-al uncharged Nariai background is \cite{b: Nariai solution} 
\begin{equation} \label{e: metric Nariai}
{\rm d} s^2 = -(1 - \sigma r^2)\, {\rm d} t^2 + \frac{ {\rm d} r^2}{(1 - \sigma r^2) } +  a^2 \,{\rm d} \Sigma_{D-2}^2 ,
\end{equation} 
where $\dd \Sigma_{D-2}^2$ denotes the line element of a $(D-2)$-dimensional unit sphere, 
\begin{equation} \label{eq: sigma a Nariai} 
 \sigma = (D-1) \Lambda, \qquad \qquad a^2  =  \frac{(D-3)}{(D-1)\Lambda},
\end{equation}
and $\Lambda$ is related to the cosmological constant. If $\sigma > 0$ then the Nariai spacetime (\ref{e: metric Nariai}) has cosmological horizons at \cite{b: Nariai solution}
\begin{equation}
 r= \pm \frac{1}{\sqrt{\sigma}}.
\end{equation} 

The QNF of the $D$-dimensional uncharged Nariai spacetime are calculated exactly in Refs.\ \cite{LopezOrtega:2007vu}, \cite{Vanzo:2004fy}, \cite{LopezOrtega:2009qc}. For the gravitational perturbations these are equal to \cite{Vanzo:2004fy,LopezOrtega:2009qc}
\begin{equation} \label{eq: Nariai QNF}
\omega = \mathbb{A} - i \sqrt{\sigma}\left( n + \frac{1}{2}\right),
\end{equation} 
where $n=0,1,2,\dots$, is the mode number, as in the previous sections. For the tensor type, vector type, and scalar type gravitational perturbations the values of the quantities $\mathbb{A}$ are given in formulas (63), (64), and (65) of Ref.\ \cite{LopezOrtega:2009qc}. In what follows we do not need the values of $\mathbb{A}$.

The Hawking temperature of the cosmological horizon for the Nariai spacetime is \cite{Vanzo:2004fy}
\begin{equation}
 T_N = \frac{\hbar \sqrt{\sigma}}{2 \pi}.
\end{equation} 
From formula (\ref{eq: Nariai QNF}) we find that for the fundamental mode of the gravitational perturbations
\begin{equation}
 \omega_I = \frac{\sqrt{\sigma}}{2},
\end{equation} 
and therefore the quantity $\mathbb{H}$ is equal to
\begin{equation}
 \mathbb{H} = 1.
\end{equation} 
Thus for the $D$-dimensional uncharged Nariai spacetime the fundamental QNF of the gravitational perturbations saturates upper bound (\ref{eq: Hod bound}), (as the fundamental mode of the extreme Kerr black hole \cite{Hod:2007tb}) and the decay time of its fundamental QNM is equal to $\tau_{min}$, in accord with the TTT bound (\ref{eq: Hod bound one}).\footnote{This fact was pointed out to me by Professor S.\ Hod in a comment on Ref.\ \cite{LopezOrtega:2009qc}. This comment and the results of Sect.\ 3 in Ref.\ \cite{LopezOrtega:2010} motivated this work. I thank to Professor Hod. }

It is convenient to comment that exists a charged generalization of the $D$-dimensional uncharged Nariai spacetime \cite{Kodama:2003kk}. Its metric is given by formula (\ref{e: metric Nariai}), but in the charged Nariai spacetime the quantities $\sigma$ and $a$ are different. The parameter $\sigma$ changes to
\begin{equation}
 \sigma = (D-1) \Lambda -\frac{(D-3)^2 Q^2}{a^{2(D-2)}},
\end{equation} 
where the quantity $Q$ is related to the electric charge of the spacetime, and the parameter $a$ is a solution to the equation
\begin{equation}
 \frac{D-3}{a^2} = (D-1) \Lambda + \frac{(D-3) Q^2}{a^{2(D-2)}} .
\end{equation}  

For the $D$-dimensional charged Nariai spacetime the QNF of the coupled electromagnetic and gravitational perturbations are calculated exactly in Appendix A of Ref.\ \cite{LopezOrtega:2007vu}. They take the same mathematical form that QNF (\ref{eq: Nariai QNF}) of the uncharged Nariai spacetime. Thus for the charged Nariai spacetime the imaginary part of the QNF depends on the mode number $n$ and the parameter $\sigma$ in the same way that the QNF of the uncharged Nariai spacetime, but the values of the quantities $\mathbb{A}$ are different. (See formulas (35), (38), and (42) of Ref.\ \cite{LopezOrtega:2007vu} for the values of the parameters $\mathbb{A}$ corresponding to the coupled perturbations of the $D$-dimensional charged Nariai spacetime.)

A similar calculation to that for the uncharged Nariai spacetime shows that for the coupled electromagnetic and gravitational perturbations of the $D$-dimensional charged Nariai spacetime, the fundamental QNF saturates upper bound (\ref{eq: Hod bound}) and therefore its decay time is $\tau_{min}$.

As in $D$-dimensional de Sitter spacetime, we notice that Hawking's temperature of the $D$-dimensional Nariai horizon depends on the inverse of the horizon radius, which is the characteristic length of the Nariai spacetime. In contrast to $D$-dimensional de Sitter spacetime the fundamental QNF of the $D$-dimensional uncharged and charged Nariai spacetimes saturates upper bound (\ref{eq: Hod bound}).

\section{Discussion}
\label{section 5}

We find that the fundamental QNF of the uncharged and charged $D$-dimensional Nariai spacetimes saturate upper bound (\ref{eq: Hod bound}) in a similar way to the extreme Kerr black hole. For the three-dimensional spinning BTZ black hole, the $D$-dimensional de Sitter spacetime, and the $D$-dimensional MTBH (see also Ref.\ \cite{LopezOrtega:2010}) we find that their fundamental QNF do not satisfy upper bound (\ref{eq: Hod bound}). Furthermore these results imply that for these gravitational systems the decay time of the fundamental QNM is less than $\tau_{min}$ of formula (\ref{eq: Hod bound one}). 

It is convenient to mention that in some papers it is stated that the TTT bound is universal, and sometimes it is called \textit{``universal relaxation bound''} \cite{Hod:2007tb}, but in our previous results we find some spacetimes for which the TTT bound is not valid, therefore it is convenient to determine when lower bound (\ref{eq: Hod bound one}) and its form (\ref{eq: Hod bound}) are valid for gravitational systems.

Although the spacetimes studied here, from a physical viewpoint are not as relevant as the Schwarzschild and Kerr black holes, the failure to satisfy upper bound (\ref{eq: Hod bound}) of the spinning BTZ black hole, the $D$-dimensional de Sitter spacetime, and the $D$-dimensional MTBH  is an interesting result, since it point outs some limitations on the applicability of bound (\ref{eq: Hod bound}) to gravitational systems. We also note that the three spacetimes not satisfying upper bound (\ref{eq: Hod bound}) are not asymptotically flat, for example, the BTZ black hole and the $D$-dimensional MTBH are asymptotically anti-de Sitter.

For the perturbed gravitational systems not satisfying upper bound (\ref{eq: Hod bound}), we think that the decay times of their QNM must be described classically. Also, these systems have well defined thermodynamic interpretations. Thus an issue deserving further study is to investigate whether our previous results imply that for the gravitational systems not satisfying upper bound (\ref{eq: Hod bound}) their thermodynamic relaxation times are larger than the decay times of their QNM \cite{Ropotenko:2007jm}. The search for additional consequences of our results deserves detailed study. Furthermore we must test the TTT bound in other gravitational systems to know whether the TTT bound is fulfilled. 

It is convenient to note that taking into account the AdS/CFT correspondence the TTT bound can be used to impose a bound on the imaginary parts of the poles for the retarded Green function of the dual CFT (see Sec.\ \ref{section 2}). We think that this fact deserves additional study.

\section{Acknowledgments}

This work was supported by CONACYT M\'exico, SNI M\'exico, EDI-IPN, COFAA-IPN, and Research Projects SIP-20100777 and SIP-20100684.

\end{document}